\documentclass[conference]{IEEEtran}
\IEEEoverridecommandlockouts
\usepackage{cite}
\usepackage{amsmath,amssymb,amsfonts}
\usepackage{algorithmic}
\usepackage{graphicx}
\usepackage{textcomp}
\usepackage{xcolor}
\usepackage{breqn}
\def\BibTeX{{\rm B\kern-.05em{\sc i\kern-.025em b}\kern-.08em
    T\kern-.1667em\lower.7ex\hbox{E}\kern-.125emX}}
\begin{document}

\title{Survival prediction and risk estimation of Glioma patients using mRNA expressions\\
}
%1\textsuperscript{st} 
\makeatletter
\newcommand{\linebreakand}{%
  \end{@IEEEauthorhalign}
  \hfill\mbox{}\par
  \mbox{}\hfill\begin{@IEEEauthorhalign}
}
\makeatother

\author{
\IEEEauthorblockN{Navodini Wijethilake}
\IEEEauthorblockA{\textit{Department of Computer Science Engineering} \\
\textit{University of Moratuwa}\\
Sri Lanka \\
navodiniw@cse.mrt.ac.lk}
% \linebreakand
\and
\IEEEauthorblockN{ Dulani Meedeniya}
\IEEEauthorblockA{\textit{Department of Computer Science Engineering} \\
\textit{University of Moratuwa}\\
Sri Lanka \\
dulanim@cse.mrt.ac.lk}
\linebreakand
\IEEEauthorblockN{ Charith Chitraranjan}
\IEEEauthorblockA{\textit{Department of Computer Science Engineering} \\
\textit{University of Moratuwa}\\
Sri Lanka \\
charithc@cse.mrt.ac.lk}
% \linebreakand
\and
\IEEEauthorblockN{Indika Perera}
\IEEEauthorblockA{\textit{Department of Computer Science Engineering} \\
\textit{University of Moratuwa}\\
Sri Lanka \\
indika@cse.mrt.ac.lk}
}

\maketitle

\begin{abstract}
Gliomas are lethal type of central nervous system tumors with a poor prognosis. Recently, with the advancements in the micro-array technologies thousands of gene expression related data of glioma patients are acquired, leading for salient analysis in many aspects. Thus, genomics are been emerged into the field of prognosis analysis. In this work, we identify survival related 7 gene signature and explore two approaches for survival prediction and risk estimation. For survival prediction, we propose a novel probabilistic programming based approach, which outperforms the existing traditional machine learning algorithms. An average 4 fold accuracy of 74\% is obtained with the proposed algorithm. Further, we construct a prognostic risk model for risk estimation of glioma patients. This model reflects the survival of glioma patients, with high risk for low survival patients. 

% Recently, many approaches are been investigated to detect gliomas, predict recurrence and also survival. 
\end{abstract}
 

\section{Introduction}
% Domain Overview  

Gliomas are the most common central nervous system tumor, that derive from neuroglial cells and progenitor cells \cite{aquilanti2018updates}. Gliomas account for 30\% of the primary brain tumors and 80\% of the malignant brain tumors, causing majority of deaths from primary brain tumors \cite{weller2015glioma}. Despite the treatment, the aggressive forms of gliomas have a mortality within months. Nowadays, mostly the treatment planning in glioma patients rely on histology and other clinical parameters such as age. Based on the histology, Gliomas are been classified into astrocytic, oligodendrogial and ependymal tumors, and additionally considering the malignancy, the natural disease cause, absense and presense of anaplastic features, they are been assigned into World Health Organization (WHO) I-IV grades \cite{fuller20072007}. Yet, the intra-tumor heterogeneity and alterations in molecular levels are associated with the prognosis of glioma patients than the underlying histology \cite{gravendeel2009intrinsic}. 

%why gene expression is important
Recently, with the progression of the  genomic, transcriptomic and epigenetic profiling and with the technological advances in Ribonucleic acid (RNA) sequencing, which is a molecule that support several biological roles of genes, and microarrays, novel approaches for classifying and analysing gliomas are recognized. Moreover, these underlying molecular pathogenesis lead to identify genetic alterations which can cause gliomas that can also be complementary to histological classifications and diagnostics \cite{weller2015glioma}. Thus, following the WHO 2016 classification, Isocitrate dehydrogenase (IDH) family IDH1/2 mutant, 1p/19q co-deleted tumors, mostly with oligodenrial histology have the best prognosis and belongs to WHO grade II. WHO grade III gliomas are IDH1/2 mutant, 1p/19q non co-deleted, Telomerase Reverse Transcriptase (TERT) promoter-wild type, tumor protein p53 (TP53) mutant tumors with an astrocytic morphology and have an intermediate survival. IDH wild-type, 1p/19q non codeleted tumors have a poor prognosis and are mostly grade IV Glioblastoma multiforme (GBM). In addition to these, there are several other molecular signatures associated with gliomas in diagnosis and prognosis, such as TERT, TP53, O-6-methylguanine Deoxyribonucleic acid (DNA) methyltransferase (MGMT) and phosphatase and tensin homolog (PTEN) \cite{ludwig2017molecular}.

%current issues and motivation
Generally oncologists assess the survival of patients, based on their experience and clinical factors. There factors include the size, location and stage of the cancer for the survival estimation. Hence, these decision can be biased, optimistic and inaccurate \cite{moghtadaei2014predicting}. However, accurate survival prediction approaches lead to less invasive better treatment with optimal usage of resources \cite{bal2015patterns}. Gene expression data is used in several types of cancers for survival prediction and shows promising improvements over other traditional approaches such as radiomic based algorithms \cite{bashiri2017improving,wijethilake2020radiogenomics,wijethilakernoai2020}. As a supervised machine learning algorithm, artificial neural networks are developed for survival prediction using gene expression data for many cancer types \cite{chen2009artificial,rasanjana2019svm}. However the use of machine learning based algorithms on gene expression data are not much seen for survival prediction in days or in class wise of glioma patients. 

In our previous work \cite{wijethilake2020radiogenomics}, we have demonstrated that incorporating both imaging biomarkers and gene expression biomarkers outperforms the survival prediction accuracy of glioblastoma patients compared to using both features separately. However, the lack of data is a limitation of that study and thus have motivated our study to have comparatively large dataset by using all the glioma cases, both higher grade and lower grade gliomas, of The Cancer Genome Atlas (TCGA) and Chinese Glioma Genome Atlas (CGGA) cohorts. Thus, we expect that this work will lead for new directions in survival predictions of glioma patients, focusing more on genomics. 

% Despite, deep learning based algorithms are been emerged in cancer classifications with an average precision of 96.25\% \cite{karim2019onconetexplainer}. Therefore, the ability of deep learning algorithms to cooperate with highly variable and high dimensional data such as gene expression data has imposed challenges to traditional bioinformatics tools. \\

%what is going to present
The proposed study has two major parts of survival estimation in glioma patients. We depict machine learning algorithms for survival prediction of glioma patients where we execute prediction after performing feature selection on high dimensional gene expression data. Further, we propose a novel probabilistic programming based Bayesian neural network for survival prediction of glioma patients in 3 categories; long, short and medium survival. The selected prognostic genes are also used to develop a prognostic risk score model. The main outline of the work is shown in Fig. \ref{fig:Flowchart}. The main objective of this study is to identify the prognostic gene signature and construct a prognostic model for risk estimation, that provide useful insights for the prognosis of glioma patients. Thus, the 7 gene signature based survival prediction algorithm with probabilistic programming and prognostic risk model are the novel contributions of this study.

% \section{Approach}

\begin{figure}[htbp]
\centering
\includegraphics[width=0.45\textwidth]{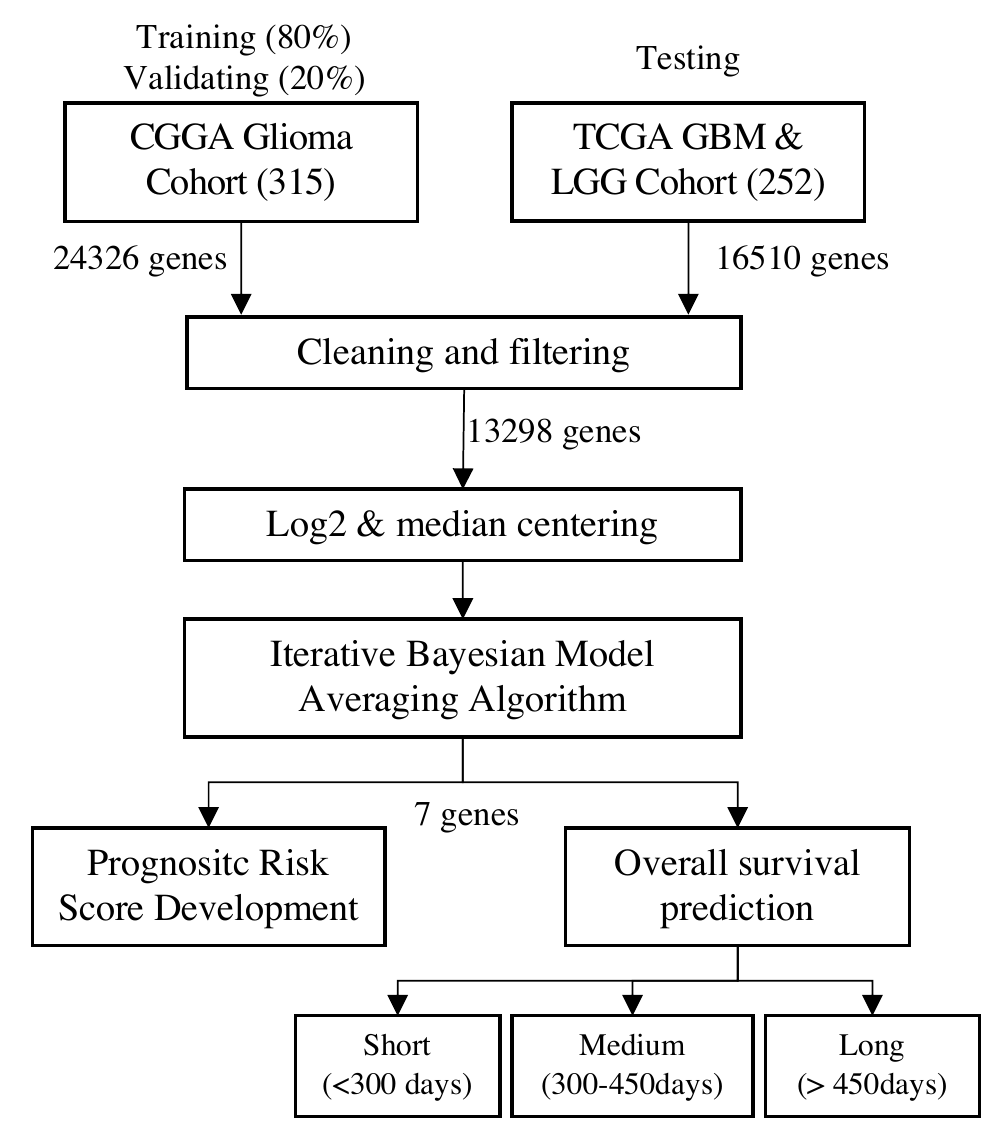}
\caption{Overview of the proposed solution.}
\label{fig:Flowchart}
\end{figure}

The paper is structured as follows: Section II explores the related work, Section III explains our approach for the analysis and Section IV reveals our observation and the evaluation summary. In Section V we discuss the important aspects, limitations of our research.

\section{Background}

The mutations, methylations and other phenomena occur in molecular level reflect the associations with the expression levels of those biomarkers \cite{labussiere2014tert}. TP53 is recognized as a negative prognostic marker, causing poor prognosis in astrocytic and oligoastrocytic gliomas. Hence, the expression of TP53 gene has an inverse correlation with survival of glioma patient \cite{wang2014gain}. Correspondingly, TERT promoter mutation is common in gliomas and also shows a negative prognosis in glioma patients \cite{labussiere2014tert}. In fact, the expression of TERT gene is significantly high with mutated TERT promoters \cite{labussiere2014tert}. Given the above, it is clear that expression of genes in gliomas are significantly associated with other molecular biomarkers and also effect the survival and prognosis of glioma patients.  

Gene expression profiling is commonly used for clustering and subtype classifications  of gliomas, using both unsupervised and supervised approaches \cite{verhaak2010integrated,vitucci2011gene}. Artificial neural network (ANN) based subtype classification is also performed on gene expression profiles of glioma patients \cite{petalidis2008improved}. 

Several studies related to cancer prognosis have used gene expression details of genetic biomarkers for predicting the cancer occurrence, for predicting the recurrence of cancers and also for predicting outcomes after the diagnosis, such as mortality, life expectancy, drug sensitivity etc. The first application of machine learning, an artificial neural network is found in early 1995 \cite{faraggi1995neural}. Moreover, O'Neill et al \cite{o2003neural} also have employed a neural network for diagnosing diffuse large B-cell lymphoma with micro-array gene expression profiles. Later, Chen et al \cite{chen2009artificial} have proposed gene expression based artificial neural network for predicting survival time of lymphoma patients. Similar methodology is employed by Lancashire et al \cite{lancashire2010validated} for breast cancer survival outcome prediction.

Nonetheless, Gene expression profiling is utilized for estimating survival of glioma patients, as it is capable of revealing the unrecognized heterogeneity of gliomas through hierarchical clustering \cite{freije2004gene}. Bonata at el \cite{bonato2011bayesian} have proposed a Bayesian ensemble model for survival prediction with high dimensional gene expression profiles by selecting the genes potentially related gliomagenesis. Identifying the potential biomarkers related to glioma survival is also typically involves gene expression profile analysis \cite{hsu2019identification}. Typically, utilizing neural networks the survival outcome of neuroblastoma patients after 5 years from the diagnosis is predicted with expression data \cite{wei2004prediction}. 

Risk score formulation, as a prognosis estimation tool, is also established with the gene expression data for Glioma patients \cite{zeng2018integrative,zuo2019rna}. For this the most prominent features are chosen with statistical methods based on the relationships of each gene and survival. The predominant statistical analysis based approaches are univariate and multivariate cox proportional hazard regression analysis. Thus, based on the Hazard ratio and the p value of each gene, the features associated with survival are identified. In some studies \cite{zeng2018integrative}, the features have chosen based on the methylation status of each gene as well. However, some studies have mentioned that risk score is not an accurate reflection of the survival probability \cite{zuo2019rna}. 

Nomograms are another initiation for estimating survival probability after a particular time period of glioblastoma patients \cite{wang20195,gittleman2020independently}.  \\

\section{System Model and Methodology}
\subsection{Dataset}
The publicly available gene expression GBM and lower-grade glioma (LGG), which are biomarkers to classify subjects into risk groups, datasets of TCGA and CGGA are downloaded for our study. TCGA dataset comprises of 252 subject cases with overall survival information and gene expression profiles of 16510 genes, obtained using the Illumina HiSeq RNA Sequencing platform. Particularly, WHO grade II, III are included in the TCGA-LGG cohort and WHO grade IV is included in the TCGA-GBM cohort.  CGGA dataset consists of 315 cases with overall survival data and gene expression profiles of 24326 genes, acquired using Illumina HiSeq 2000 platform, which is a powerful high-throughput sequencing system.  The initial gene expression values, are normalized based on the gene length into fragments per kilo-base per million mapped reads (FPKM) \cite{abbas2018comparison} in comma-separated values (CSV) format. 

\begin{equation}
\mathrm{FPKM}=\frac{\text {Total fragments
mapped reads million}}{\text {Exon length kilobase pair} }
\end{equation}
Moreover, all the gene expression features are log transformed and normalized median centered before analysing. After normalization, 13094 gene common for both TCGA and CGGA datasets are chosen. The CGGA dataset is divided into training and validating datasets while the TCGA dataset is used for testing. Based on the overall survival in days, the classes of the each patient are determined, where the three classes are, short survival (overall survival in days $<$ 300 days), medium survival (300-450 days of survival) and long survival (overall survival $>$ 450 days). The dataset distribution for the classes short, medium and long survival are given in Table \ref{tab1:distribution}. The CGGA dataset is divided in 4 folds, with equal distribution of classes in each fold and the ratio of classes in training and validating folds are maintained without overlapping cases. Table \ref{tab1:distribution} shows the class distribution in each dataset, CGGA and TCGA cohorts and in the training and validating datasets of CGGA cohort. 

\begin{table}[htbp]
\caption{Dataset Description}
\begin{center}
\begin{tabular}{|c|c|c|c|}
\hline
\textbf{Class}&\multicolumn{2}{|c|}{\textbf{CGGA dataset}} & \textbf{TCGA} \\
\cline{2-4} 
\textbf{} & \textbf{\textit{Training}}& \textbf{\textit{Validating}}& \textbf{\textit{Testing}} \\
\hline
Short & 63&21 & 68 \\ \hline
Medium & 29 & 9& 44\\ \hline
Long & 145 &48 & 140\\ \hline
\end{tabular}
\label{tab1:distribution}
\end{center}
\end{table}

\subsection{Prognostic Gene Identification}
We utilize Bayesian Model Averaging (BMA) \cite{annest2009iterative} based feature selection approach in order to obtain a robust learning model. BMA algorithm overcomes the model uncertainty by obtaining the average over the posterior probability distributions of several models. Thus, the posterior probability of being $\Phi$ given the training dataset $D$ can be as follows.

\begin{equation}
\operatorname{Pr}(\Phi | D)=\sum_{i \in S} \operatorname{Pr}\left(\Phi | D, M_{i}\right) \cdot \operatorname{Pr}\left(M_{i} | D\right)
\end{equation}

where, $M_{i}$ is a given model in the subset, $S=\left\{M_{1}, M_{2}, \ldots, M_{n}\right\}$.
Initially the genes are ranked based on Cox proportional hazard analysis (Cox-PH), which has the ability to deal with censored data. Cox hazard function (\eqref{equation:cox-hazard}) for a given covariate vector of subject $p$, $\mathbf{z}_{p}=\left(z_{1 p}, \dots, z_{i p}\right)$, depicts the probability of dying at a given time $T$, if the patient survived until time $T$.

\begin{equation}
\lambda\left(\mathrm{T} ,\mathbf{z_{p}}\right)=\lambda_{0}(\mathrm{T}) \exp \left(\mathbf{z_{p}} \alpha\right)
\label{equation:cox-hazard}
\end{equation}

where baseline hazard function is $\lambda_{0}$ and the coefficients of the each $\left(z_{1 p}, \dots, z_{i p}\right)$ covariates is given by $\alpha = \left(\alpha_{1 p}, \dots, \alpha_{i p}\right)$. Since the baseline hazard function is same for a single case, for different time $T$, it can be neglected. Thus, an approximation for $\alpha$ is required and it can be calculated with the following partial likelihood association. 

\begin{equation}
\operatorname{PL}(\alpha)=\prod_{p=1}^{n}\left(\frac{\exp \left(\mathbf{z_{p}} \alpha\right)}{\sum_{\ell \in R_{p}} \exp \left(\mathbf{z_{\ell}} \alpha\right)}\right)^{\delta_{x}}
\label{equation:partial}
\end{equation}

In this Equation \eqref{equation:partial}, $R_{p}$ is the risk set, subjects that have not experienced an event by the time $t_{p}$ and $\delta_{i}$ is the event status (censored or not) of patient $x$. The $\theta$ parameter is obtained by maximizing the partial likelihood, and thus, according descending order of log likelihood of those values, genes are ranked.

The top ranked 25 genes are assigned to a window and traditional BMA algorithm is applied for survival analysis. Based on the posterior probabilities of those genes, the genes with low posterior probabilities ($<1\%$) are eliminated retaining the genes with high posterior probabilities. The window is moved along the top ranked genes until 7 genes with high posterior probabilities are obtained. These 7 genes and their posterior probabilities are given in the Table \ref{tab:posterior_prob}. Accordingly, TXLNA gene has the highest posterior probability and the other genes have posterior probabilities between 1 and 100\%. 

\begin{table}[htbp]
\caption{Selected 7 genes and their corresponding posterior probabilities}
\begin{center}
\begin{tabular}{|c|c|}
\hline
\textbf{Gene}&\textbf{Posterior probability} \\ \hline
TXLNA  & 100.0 \\ \hline
 WDR77 & 50.5  \\ \hline
 TAF12 & 48.9  \\ \hline
STK40  & 35.9  \\ \hline
YTHDF2 & 14.7. \\ \hline
SNRNP40  & 8.0.  \\ \hline
SLC30A7& 2.0   \\ \hline
\end{tabular}
\label{tab:posterior_prob}
\end{center}
\end{table}

\subsection{Survival Prediction}

We utilized the most commonly used machine learning algorithms to predict the overall survival class, short, medium and long. For all these the input is the selected 7 gene signature. In fact, To overcome the class imbalance, shown in Table \ref{tab1:distribution}, the minority class (i.e. short and medium classes) entries in the training dataset are over sampled randomly.

\subsubsection{Decision Tree}

Decision tree (DT) \cite{marubini1983prognostic}  is a tree based structure, where each node in the tree specifies a condition of a input covariate that decides the distribution or the final class. The decision tree is fine tuned to obtain the best results, with a maximum depth of the tree to be 4 and the minimum samples required to make a decision at a node to be 10.  

\subsubsection{Random Forest Classification} 

Random Forest Classifier (RFC) \cite{breiman2001random} is a widely used ensemble model, that outperforms a single decision tree by overcoming the instability and the variance, with multiple decision trees. Hence, in RFC the number of trees in the forest are tuned to 100.
\subsubsection{XGBoost} 

XGBoost \cite{chen2016xgboost} is an ensemble model but each tree is trained in a sequential manner, boosting the prediction accuracy. The fine tuned XGBoost model, consists of 50 trees trained sequentially, with a maximum depth of a tree, 4 layers. 

\subsubsection{CatBoost}

Catboost \cite{prokhorenkova2018catboost} is a sequentially trained boosting algorithm, which outperforms the existing boosting algorithms with its efficiency. We utilize multi-class loss function with a learning rate of 0.005 for 2000 epochs for 4 layered symmetric tree. 

\subsubsection{Support Vector Machine} 

Support Vector Machine (SVM) \cite{smola2004tutorial} is a common machine learning algorithm that is been used for survival prediction of glioma patients. Thus, we explore a support vector classification based on SVM with a linear kernel.

Moreover, we propose a novel Bayesian neural network based on deep probabilistic programming. Deep probabilistic programming uses the power of deep neural networks with probabilistic models, to enhance the performance while optimizing the related costs \cite{ rubasinghe2019dppl}. The traditional more frequently utilized algorithms show less performance, as discussed in the results section. Therefore, a need for a better performing algorithm, which can recognize the uncertainties in the predictions, occurred. 

\subsubsection{Proposed  Bayesian Neural Network}

We propose probabilistic programming based Bayesian neural network (BNN) for the survival class prediction of glioma patients. The first layer receives the chosen 7 features as the input. The final layers contains the predicted class, short, medium or long. In the middle the architecture consists of 2 layers comprised with 24 and 12 neurons in each layer respectively. Both hidden layers are comprise of 50\% dropout with $1^{st}$ hidden layer followed by a Tanh activation and the $2^{nd}$ hidden layer followed by a rectified linear activation unit (ReLU). All the nodes are fully connected with nodes in the adjacent layers. BNN differs from the traditional artificial neural network by assigning distributions for the all the parameters, weights and biases instead of a single value. Mean and scale parameters of all the distributions are initialized to 0.01 and 0.1 respectively. Stochastic gradient descent optimization with learning rate 0.001 is utilized for training the BNN. Further, stochastic variational inference optimizes the trace implementation of evidence lower bound (ELBO) in order to diverge probability distribution of parameters. The proposed BNN architecture is shown in Fig. \ref{fig:BNN}.

Implementations are developed with Pyro (version 1.3.1) probabilistic programming language with pytorch (version 1.5.0) \cite{bingham2019pyro}. For the other machine learning algorithms Scikit-learn library \cite{scikit-learn} is utilized. To measure and compare the performance of the applied machine learning algorithms, 4 metrics Accuracy, Precision, Recall and F1 scores are occupied. 
\begin{figure}[ht!]
\centering
\includegraphics[width=0.5\textwidth]{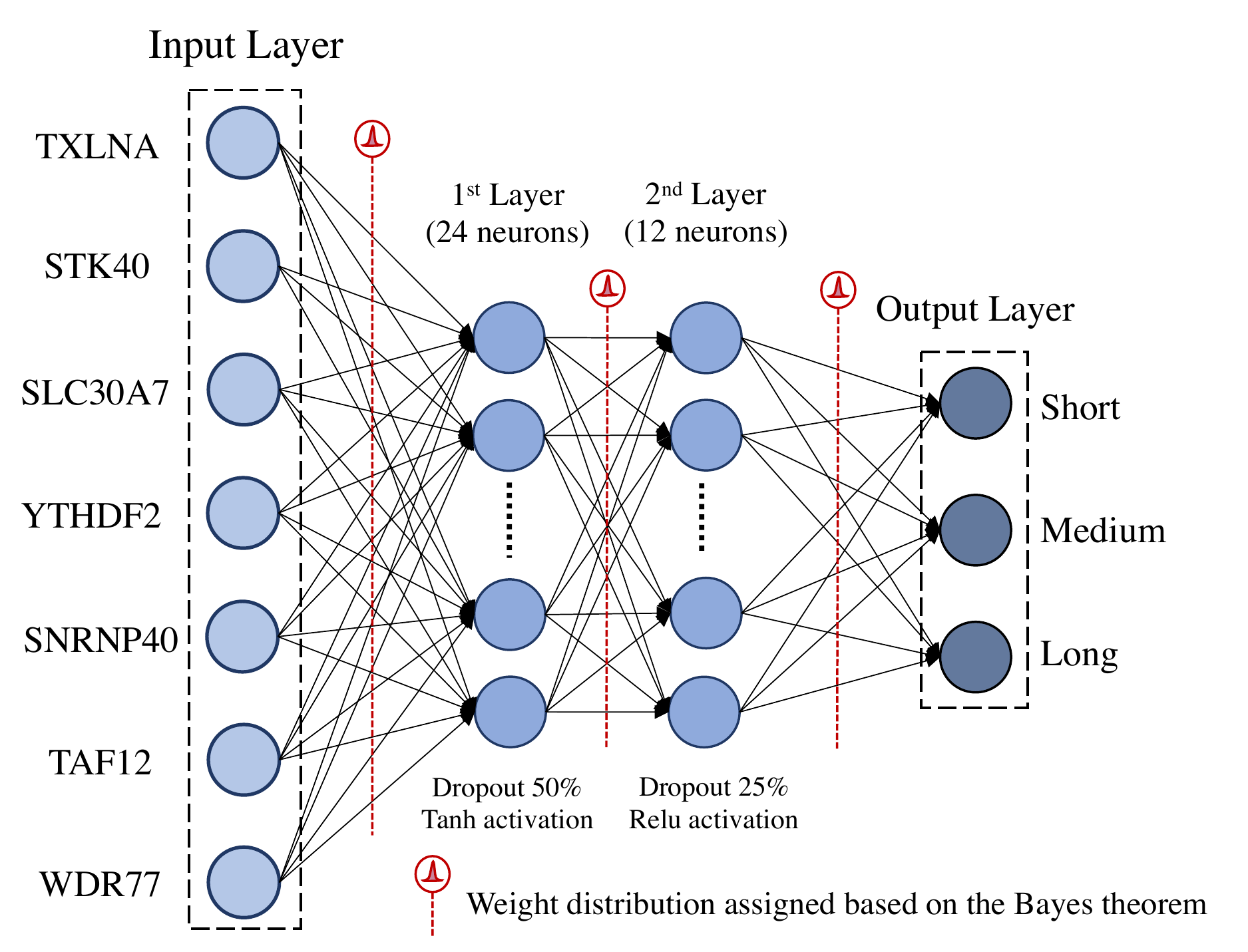}
\caption{Proposed Bayesian Neural network architecture.}
\label{fig:BNN}
\end{figure}

 \subsection{Prognostic Risk model Construction}
Further, in order to estimate a risk, we calculate a risk score with the survival related genes of glioma patients.  We use univariate cox proportional hazards regression analysis to evaluate the association between the 7 genes obtained from the iterative Bayesian Model Averaging algorithm and the survival status and time. Typically the genes with p values $< 0.01 $ and the hazard ratio (HR) $>1$ are considered as the criteria for them to be candidate genes for the survival estimation. The obtained $\beta$ values, i.e. the regression coefficient obtained from the cox analysis are used to calculate the risk score along with the expression values of the corresponding genes ($\exp_{i}$). The univariation cox analysis is performed with the R package, survival (version 3.2-3) \cite{survival-package}. The Equation \eqref{eq:risk_score} indicates the prognostic risk score formula used to obtain the risk score prognostic model. 
\begin{equation}
\text {Risk score}=\sum_{i=1}^{n} \beta_{i} * \exp _{i}
\label{eq:risk_score}
\end{equation}
$n$ is the number of genes chosen to be included in the prognostic signature. After obtaining the risk score for the CGGA cohort, the median value of the prognostic risk score is used to separate the patients in the high and low risk groups. To clarify, the performance of the proposed signature is validated on the TCGA cohort.

\section{System Evaluation}
\subsection{Overall survival prediction}
The Accuracy, Precision, Recall and F1 scores were calculated to compare the techniques utilized to predict the survival class. Each algorithm were trained on CGGA training splits and validated on the corresponding fold validation split. Table \ref{tab1:ML} shows the average metric results over 4 folds CGGA validation splits. Thus, Based on the all 4 metrics the proposed BNN algorithm outperforms the rest of frequently utilized ML techniques, with 74.50\% average accuracy. According to the studies, the highest reported accuracy is 68\% for radiomics \cite{shboul2019feature}, with a cohort $>$ 100 patients used for the training. According to our results, second best performing algorithm was Random Forest Regression with over 60\% accuracy. 
 
\begin{table}[htbp]
\caption{Comparison of Overall Survival Prediction with Machine Learning - 4 fold cross validation on CGGA cohort}
\begin{center}
\begin{tabular}{|c|c|c|c|c|}
\hline
\textbf{ML methods}&  \textbf{Accuracy}& \textbf{Precision} & \textbf{Recall} & \textbf{F1 score}  \\\hline
DT \cite{marubini1983prognostic}	& 55.25\%&	57.50\%	&55.25\%&	56.00\%\\\hline
RFC \cite{breiman2001random}	&62.25\%&	61.25\%	&62.25\%&	61.00\%\\\hline
XGBoost \cite{chen2016xgboost} &56.25\%	&62.75\%&	56.25\%	&58.25\%\\\hline
CatBoost \cite{prokhorenkova2018catboost}&	57.25\%&	62.75\%&	57.25\%	&59.00\%\\\hline
SVC \cite{smola2004tutorial}	&52.75\%&	67.25\%	&52.75\%	&56.50\%\\\hline
\textbf{BNN} & \textbf{74.50\%} &	\textbf{67.50\%} &	\textbf{73.00\%}	& 70.25\% \\ \hline
\end{tabular}
\label{tab1:ML}
\end{center}
\end{table}

Further, we evaluated all the trained algorithms of each fold on TCGA cohort. Thus, we obtained the average over the metrics obtained from the testing on each fold trained with CGGA. We could clearly observe the BNN performing better compared to the other algorithms in testing phase. The support vector machine and decision trees, that are been widely utilized in survival prediction, showed a comparatively low performance with gene expression data. The Boosting algorithms showed an average performance and the ensemble algorithm, RFC indicated the second best performance with regard to the other algorithms. The performance comparison on TCGA cohort validation is given in Table \ref{tab1:ML-TCGA}.

\begin{table}[htbp]
\caption{Comparison of Overall Survival Prediction with Machine Learning - testing on TCGA cohort}
\begin{center}
\begin{tabular}{|c|c|c|c|c|}
\hline
\textbf{ML methods}&  \textbf{Accuracy}& \textbf{Precision} & \textbf{Recall} & \textbf{F1 score}  \\\hline
DT	& 44.50\%	& 46.00\%	&44.50\%	& 45.00\% \\\hline
RFC	&54.25\%	&51.25\% &	54.25\%&	51.00\% \\\hline
XGBoost	&51.00\%	& 52.25\% &	51.00\%	& 51.75\% \\\hline
CatBoost&50.25\%	& 52.75\%	& 50.25\%	& 50.75\%\\\hline
SVC	&47.25\%	& 56.00\%	& 47.25\%	&50.00\%\\\hline
BNN &\textbf{59.75\%}&	\textbf{57.25\%}&	55.25\%&	51.00\% \\ \hline
\end{tabular}
\label{tab1:ML-TCGA}
\end{center}
\end{table}

\subsection{Prognostic risk model}
We initiated a prognostic model, based on the univariate cox regression analysis coeefcients and the expression of the most prominent 7 genes, TXLNA, STK40, SLC30A7, YTHDF2, SNRNP40, TAF12 and WDR77 associated with the survival. Univariate cox regression analysis results are shown in the Table \ref{tab1:univariate}. 

\begin{table}[ht!]
\caption{Univariate Cox Regression analysis on the chosen 7 genes}
\begin{center}
\begin{tabular}{|c|c|c|c|}
\hline
\textbf{Gene}&  \textbf{Coef in coxPH}& \textbf{Hazard Ratio (95\% CI)} & \textbf{p value}   \\
\hline
TXLNA & 0.87 & 2.4 (2.1-2.7) & 2.2$e^{-36}$  \\ \hline
STK40 & 0.77 & 2.2 (1.9-2.4) & 4.8$e^{-33}$  \\ \hline
SLC30A7 & 0.8 & 2.2 (2-2.5) & 2.6$e^{-33}$  \\ \hline
YTHDF2 & 0.74 & 2.1 (1.8-2.4) & 1.5$e^{-29}$  \\ \hline
SNRNP40 & 0.73 & 2.1 (1.8-2.4) & 1.2$e^{-31}$  \\ \hline
TAF12 & 0.73 & 2.1 (1.8-2.3) & 1.5$e^{-34}$  \\ \hline
WDR77 & 0.77 & 2.2 (1.9-2.5) & 5.1$e^{-32}$  \\ \hline
\end{tabular}
\label{tab1:univariate}
\end{center}
\end{table}

This analysis proved that the genes chosen with the iterative BMA algorithm are highly associated with the survival, with HR over 2 for all the 7 genes. All the coefficients of cox analysis were statistically significant (p values $<$ 0.01 for all the 7 genes) and all the coefficients were positive. Accordingly we can conclude that the high expression of these genes are associated with the poor survival of glioma patients. Thus, the risk score formula is shown in Equation (6).

\begin{dmath}
\text{Risk score} = 0.87 * \exp_{TXLNA} + 0.77 * \exp_{STK40} + 0.8 * \exp_{SLC30A7} + 0.74 * \exp_{YTHDF2}+ 0.73 * \exp_{SNRNP40} + 0.73 * \exp_{TAF12} + 0.77 * \exp_{WDR77}
\label{eq:prog_sig}
\end{dmath}

Based on the above prognostic risk model, the risk score was calculated for all the cases in the CGGA and TCGA cohorts. Further, the threshold of the high risk \& low risk determination, was considered as the median of the risk scores of the CGGA cohort. %Thus, the median risk score was found as $0.6396$. 

As shown in Fig. \ref{fig:heatmap}(a), it can be seen that the high risk patients have a high expression in all the 7 genes and the survival was mostly short and medium. In fact, when the 7 genes are low expressed, the majority of the patients have a long survival and thus, a low risk. 

\begin{figure}[ht!]
\centering
\includegraphics[width=0.45\textwidth]{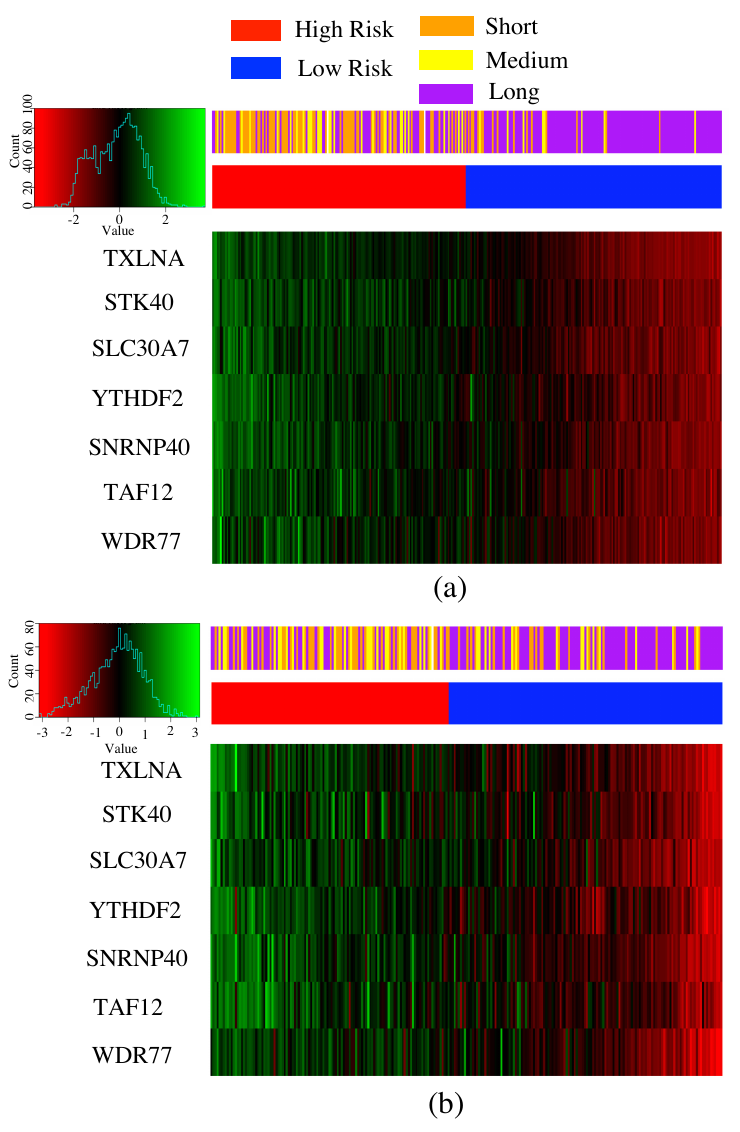}
\caption{Heat map for the (a) CGGA cohort (b) validation TCGA cohort with proposed gene signature.}
\label{fig:heatmap}
\end{figure}

This prognostic risk model was validated on the TCGA cohort by obtaining the risk score using the Equation \eqref{eq:prog_sig}. The same threshold was used to divide the patients into high and low risk groups. Consequently, the aforementioned relationship could also be seen when we observe the Fig \ref{fig:heatmap} (b) obtained for the TCGA cohort. Most of the patients with a highly expressed 7 genes also had a low survival and a high risk. 

This could further be clarified by observing the distribution of the overall survival in days in each risk group shown in Fig. \ref{fig:OSdis}. Patients with a high risk significantly had a low survival and patients with high risk had a large span of overall survival in days.

\begin{figure}[ht!]
\centering
\includegraphics[width=0.45\textwidth]{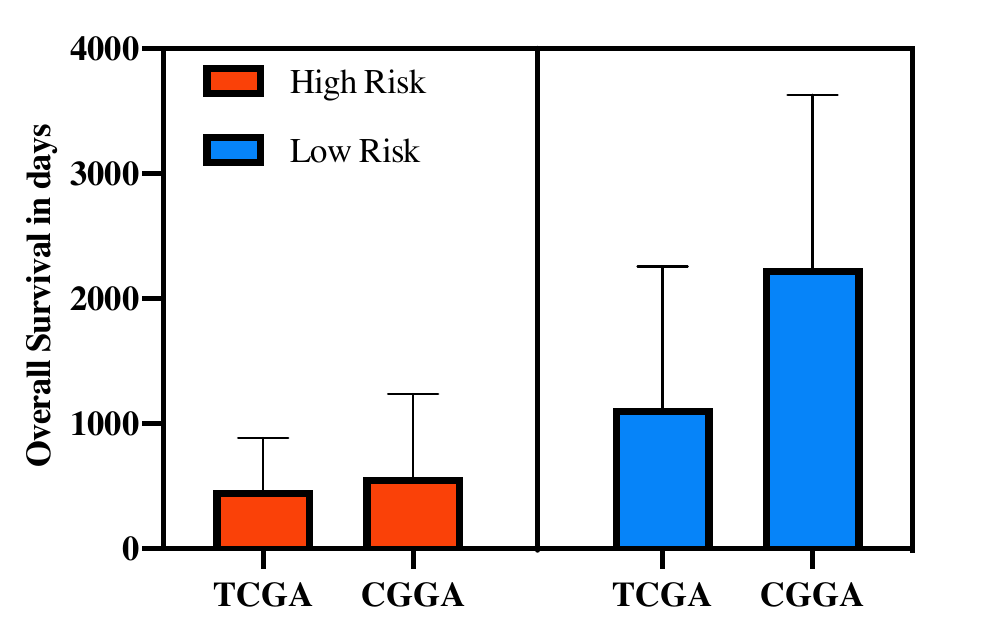}
\caption{Overall survival distribution of high and low risk groups.}
\label{fig:OSdis}
\end{figure}

For CGGA cohort for the high risk and low risk groups have an overall survival of $572.2658 \pm 662.5993$ days and $2244.361 \pm 1385.437$ days, respectively. For testing, on TCGA, the overall survival of the high risk and low risk groups were $468.1624 \pm 412.3076$ days and $1124.43 \pm 1131.1$ days, respectively. Thus, we could observe that the high risk patient group has a low survival in contrast to the low risk patient group.

Kaplan-Meier plots were acquired for both TCGA and CGGA cohorts, thus verifying the low percentage of survival with respect to the overall survival for high risk patients. Fig. \ref{fig:Kap-Mei}  verifies the performance of the prognostic risk model demonstrating the above observations. Correspondingly, The high risk group TCGA cohort demonstrated a low percentage of survival with respect to the overall survival, as shown in Fig. \ref{fig:Kap-Mei} (b). This verified the prognostic risk model behaviour for CGGA cohort, shown in \ref{fig:Kap-Mei} (a). 

\begin{figure}[htbp]
\centering
\includegraphics[width=0.45\textwidth]{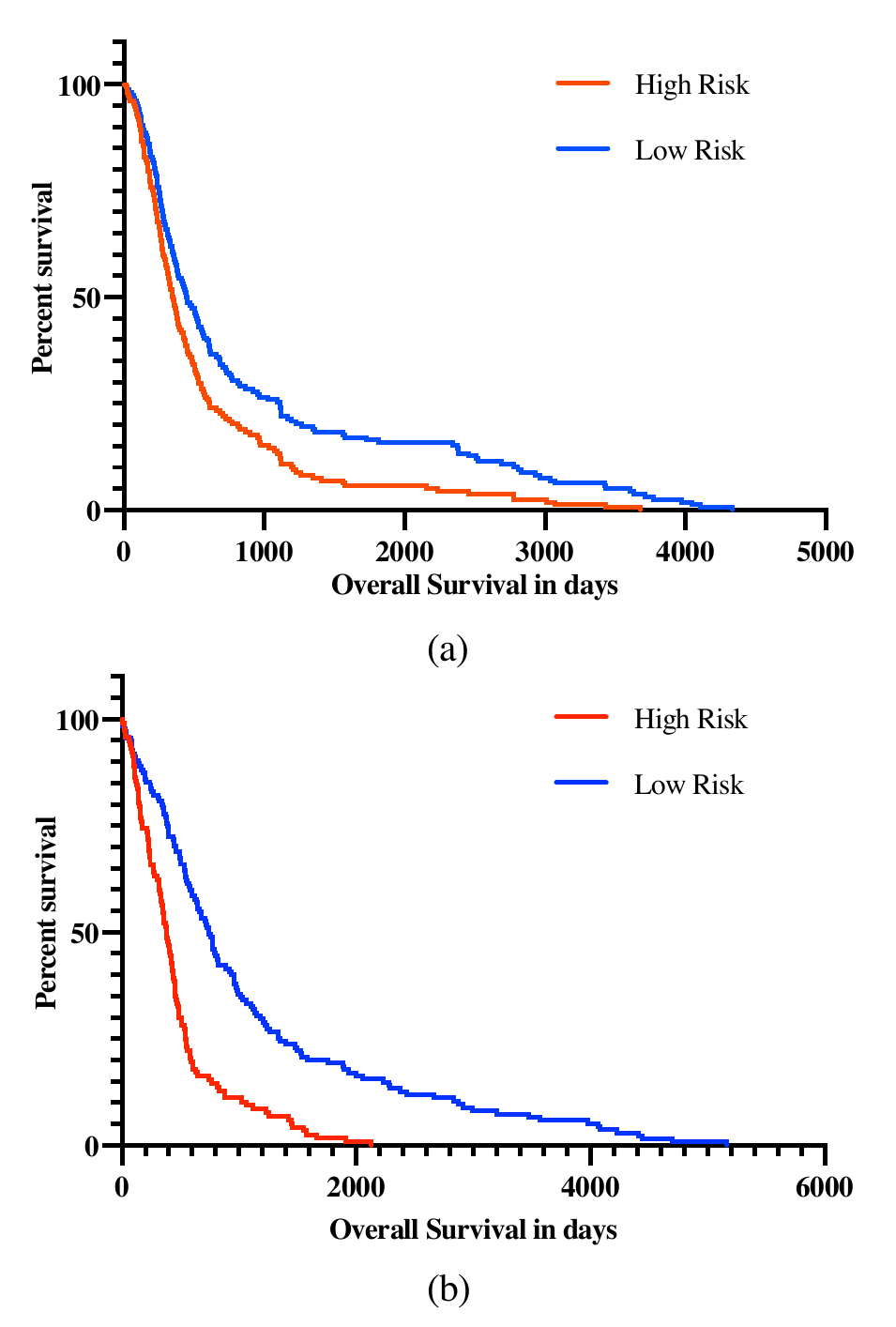}
\caption{Kaplan-Meier curves obtained for the high risk and low risk groups. (a) CGGA (b) TCGA dataset.}
\label{fig:Kap-Mei}
\end{figure}

Fig. \ref{fig:expdis} shows the gene expression distribution of the the most prominent genes we choose for prognostic risk model development, in high risk and low risk groups. It can be observed that the mean expression value of each gene in the low risk group is lower than the mean expression value of the corresponding gene in the high risk group.

\begin{figure}[htbp]
\centering
\includegraphics[width=0.5\textwidth]{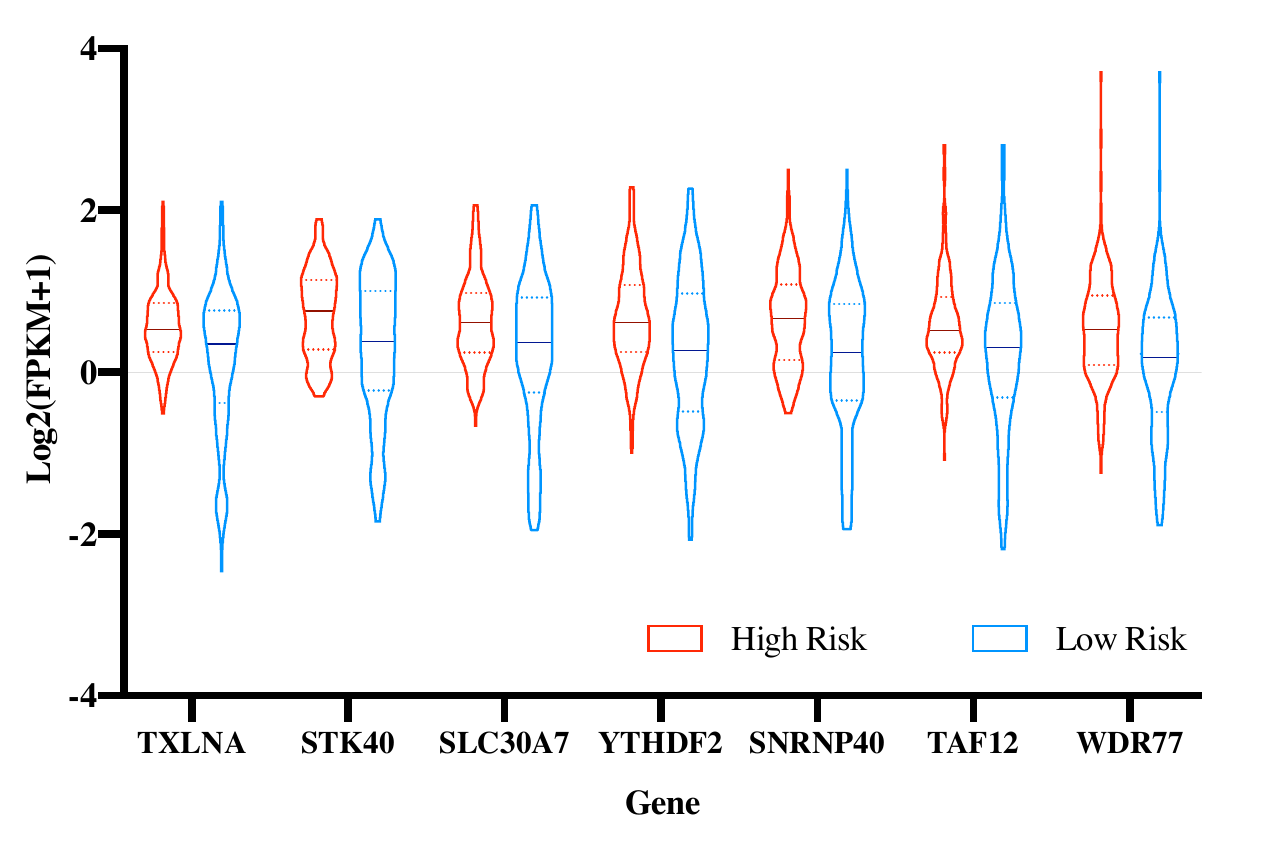}
\caption{mRNA expression value distribution of each selected genes for the high and low risk groups.}
\label{fig:expdis}
\end{figure}

% \begin{figure*}[htbp]
% \centering
% \includegraphics[width=1\textwidth]{tree.png}
% \caption{Example of a figure caption.}
% \label{fig}
% \end{figure*}

% \begin{figure}[htbp]
% \centering
% \includegraphics[width=0.45\textwidth]{CGGA.pdf}
% \label{fig:CGGA}
% \end{figure}

\section{Discussion}

Many studies reveal that genetic alterations and the molecular heterogeneity plays a vital role in gliomagenesis and prognosis. Therefore, recently many prognosis assessment tools are originated to with these different types of omics data related to gliomas. In this current study we identified 7 molecular biomarkers, with the potential of survival prediction, which depict the underlying heterogeneity. The state-of-art survival prediction in gliomas are mainly with radiomic features, extracted from tumor region \cite{3Dunet2020Islam,shboul2019feature}. As in the Brain Tumor Segmentation challenge 2018 the highest accuracy obtained is 68\% with imaging features \cite{shboul2019feature}. Here, we proposed a machine learning based approach for survival prediction with gene expression data with an accuracy over 70\%. Moreover, we established a prognostic model to estimate the risk of glioma patients. The high-throughput genomics depict underlying molecular biology of gliomas and thus, have shown improvements over radiomics for survival prediction. 

Out of the considered 7 genes namely TXLNA, STK40, SLC30A7, YTHDF2, SNRNP40, TAF12 and WDR77, the gene TAF12 is said to have associations with gliomas according to the previous studies. Mostly the grade II \& III gliomas have mutated IDH1/IDH2. As claimed by Ren et al \cite{ren2015identifying} mutated IDH1 down regulates the TAF12 expression. Our observation of low expression of TAF12 in high risk category with low survival as shown in Fig. \ref{fig:heatmap}, is on par with above findings. Yet, our findings disclose that there are 6 other genes, which has not been identified to have association with survival exists, and can be used for risk and survival estimation.

However most of these studies rely on a single cohort to avoid the inconsistencies occur between different datasets. These inconsistencies occur often due to the technical limitations, as they have been acquired from different high-throughput platforms. This can also be affected by the probes' cross hybridization, redundancy and annotations \cite{zhao2014comparison}. Therefore, the significantly low performance of TCGA compared to the CGGA in both of our survival prediction, is mostly due to these reasons. Besides, the prognostic model, as shown in fig \ref{fig:heatmap} (b), reveals that although it is developed on CGGA cohort, it still has the ability to separate patients in TCGA cohort, in to risk groups as a relatively promising prognostic model. 

For the prognostic risk score model, unlike previous studies, we consider the complete cohort of gliomas, without separating them into grades. Hence, without the prior diagnosis of the grade, the prognostic risk model can be deployed for risk estimation. Further, for this prognostic risk score model, only require 7 genes, making it less complex for the clinicians. 

According to studies, probabilistic programming languages are capable of learning with less number of samples \cite{fei2006one}. Thus, in our application with a deficit of learning samples this algorithm showed improvements over other traditional machine learning algorithms. Further, Deep Probabilistic Programming Languages (DPPL) that combines deep learning model with probabilistic programming, has shown strong potential to achieve promising outcomes in deep learning computations \cite{ rubasinghe2019dppl}. In the future, the overall survival prediction task will be extended using an explainable method in order to identify the contribution of expression of genes with a high accuracy.

\section{Conclusion}
%solution addressed
This study has presented a solution for survival prediction of glioma patients based on genomics with a comparatively large dataset. We identified 7 gene signature associated with survival and proposed two approaches for prognosis prediction which have a potential ability to separate glioma patients into groups based on their survival and risk. %aims covered 
We proposed a novel Bayesian neural network, for survival prediction that surpasses the state-of-art survival prediction approaches. Moreover, we established a comprehensive prognostic risk model based on Cox-PH, that estimates the risk of glioma patients including both GBM and LGG.  %methodology followed
Both of these approaches have promising predictive ability of survival and risk for glioma patients, with over 70\% accuracy for overall survival class prediction. %results obtained

\section{Acknowledgement}
We acknowledge the support from the Senate Research Committee Grant SRC/LT/2019/18, University of Moratuwa, Sri Lanka

\bibliographystyle{IEEEtran}
\bibliography{mybib}

\end{document}